\documentstyle{mn}
\input{epsf}

\def\fun#1#2{\lower3.6pt\vbox{\baselineskip0pt\lineskip.9pt
\ialign{$\mathsurround=0pt#1\hfill##\hfil$\crcr#2\crcr\sim\crcr}}}

\def\beq{\begin{equation}}
\def\eeq{\end{equation}}

\def\mpc {h^{-1} {\rm Mpc}}

\begin{document}
\journal{OUTP-00-33-P}
\title[The 3-point function in the large scale structure: I.
The weakly non-linear regime in N-body simulations.]
{The 3-point function in the large scale structure: I.\\
The weakly non-linear regime in N-body simulations.}
\author[J. Barriga \& E. Gazta\~naga]
 {J. Barriga$^1$ \& E. Gazta\~naga$^{1,2,3}$ \\
$^{1}$ Institut d'Estudis Espacials de Catalunya (IEEC), ICE/CSIC, 
 Edf. Nexus-201 - c/ Gran Capit\`a 2-4, 08034 Barcelona, Spain \\
$^{2}$ INAOE, Astrofisica, Tonantzintla, Apdo Postal 216 y 51, 
 Puebla 7200, Mexico \\
$^3$ Departament de Fisica Fonamental, Universitat de
Barcelona, Diagonal 647, 08028 Barcelona, Spain}
\date{\today}
\pubyear{2001}
\maketitle
\begin{abstract}
  This paper presents a comparison of the predictions for the 2 and 3-point
  correlation functions of density fluctuations, $\xi$ and $\zeta$, in
  gravitational perturbation theory (PT) against large Cold Dark Matter (CDM)
  simulations. This comparison is made possible for the first time on large
  weakly non-linear scales ($>10 \mpc$) thanks to the development of a new
  algorithm to estimate correlation functions for millions of points in only a
  few minutes.  Previous studies in the literature comparing the PT
  predictions of the 3-point statistics to simulations have focused mostly on
  Fourier space, angular space or smoothed fields.  Results in configuration
  space, such as the ones presented here, were limited to small scales 
  were leading order PT gives a poor approximation. Here we also propose and
  apply a method for separating the first and subsequent orders contributions
  to PT by combining different output times from the evolved simulations.  We
  find that in all cases there is a regime were simulations do reproduce the
  leading order (tree-level) predictions of PT for the reduced 3-point
  function $Q_3 \sim \zeta/\xi^2$.  For steeply decreasing correlations (such
  as the standard CDM model) deviations from the tree-level results are
  important even at relatively large scales, $\simeq 20$ Mpc/h. On larger
  scales $\xi$ goes to zero and the results are dominated by sampling errors.
  In more realistic models (like the $\Lambda CDM$ cosmology) deviations from
  the leading order PT become important at smaller scales $r \simeq 10$ Mpc/h,
  although this depends on the particular 3-point configuration. We 
  characterise
  the range of validity of this agreement and show the behaviour of the next
  order (1-loop) corrections.

\end{abstract}

\section{Introduction}
\label{intro}

The systematic use of the 2-point and 3-point density correlation functions in
the context of the large scale structure in the universe was introduced by
Peebles and his collaborators in the beginning of the 70's
~\cite{PB73,HP73,PH74,PB74,PG75}.  It is well known that a Gaussian field
produces a zero 3-point correlation function. 
In the standard cosmological model, the simplest (1-field) inflationary
scenarios find a Gaussian (or almost Gaussian) form for the distribution of
initial matter density and fluctuations ~\cite{F93,Ga94,G94,WK00,GM00}.
However, there are more complex inflationary models
~\cite{A87,KP88,Sa89,LM97,Pa99}, and also models based on topological defects
~\cite{V85,Va86,Hi89,T89,AS92,GM96} which predict non-gaussian initial
conditions for primordial perturbations.

Even if we start from gaussian initial conditions, gravity makes fluctuations
evolve into a non-gaussian distribution \cite{BCGS01} as will also be shown in
section ~\ref{the2}.  It is therefore important to study if there is the
possibility of disentangling the non-gaussian effects produced by gravity from
the non-gaussian characteristics of initial conditions. We would also like to
use the 3-point function to differentiate among different cosmological models
through their distinctive (non-linear) gravitational evolution. As mentioned
above, during the last $25$ years there have been many works related to the
study of the 3-point correlation function.  Most of the recent studies, have
centered on the projected distribution ~\cite{GF99,BK00} or its Fourier
counterpart, the bispectrum ~\cite{SF96,S97,S00}. In this work we deal with
the study of how to compute the 3-point correlation function from simulations
in real 3-D (or configuration) space and its comparison to PT. This topic has
been investigated before by ~\cite{MS94,JB97} but, due to difficulties in
dealing with such a large number of particles of the simulations, the study was
only done with random subsamples and only at small scales. It remains to be
seen if this is the reason why in these previous attempts it was found that
N-body simulations do not reproduce well the leading order PT predictions.

In this work we develop a means of dealing with such large numbers, explained
in section ~\ref{thealgo}, and also a method to extract non-linearities from
simulations in order to compare the result more properly to perturbation
theory values, in section ~\ref{disen}.  The actual comparison with
simulations is presented in section \S4.

\section{The 2 and 3-point function}
\label{the2}

In this section we recall the definitions of the 2 and 3-point functions. We
also revised the predictions for these quantities according to gravitational
evolution in perturbation theory starting from gaussian initial conditions.

\subsection{Definitions}

The
2-point correlation function $\xi(r_{12})$ is defined by the joint 
probability of finding one galaxy (or a point) 
in a small volume $\delta V_1$ and another one
 in the volume $\delta V_2$ separated by a distance $\vec r_{12}$, when
 choosing the two volumes randomly within a large (that is, representative) 
volume of the universe,
\begin{equation}
\delta P_2=\bar\rho^2 \left[1+\xi(r_{12})\right]\delta V_1\delta V_2
\label{firsteq}
\end{equation}
\noindent 
where $\bar\rho$ is the mean number of galaxies per unit volume. In other words, 
$\xi(r_{12})$ represents the excess probability, compared to a random 
distribution, of finding a galaxy at a distance $r_{12}$ from another galaxy.
When we approximate the distribution of objects by a continuous density 
function $\rho(\vec r)\equiv \bar\rho [ 1 +\delta(\vec r)]$ we can write
$\xi$ in terms of density fluctuations:
\begin{equation}
\xi(r_{12})=<\delta (\vec r_1)\delta (\vec r_2)>
\label{secondeq}
\end{equation}
\noindent
Due to homogeneity and isotropy $\xi$ depends only on the 
modulus of vector $\vec r_{12}$. For the 3-point correlation function we have,
 in an analogous way,
\begin{eqnarray}
\lefteqn{\delta P_3=\bar\rho^3 \left[
    1+\xi(r_{12})+\xi(r_{23})+\xi(r_{13})+\zeta(r_{12},r_{23},r_{13}) \right]
\nonumber } \\
\lefteqn{\quad \delta V_1\delta V_2\delta V_3}
\label{thirdeq}
\end{eqnarray}
\noindent 
where $ \zeta(r_{12},r_{23},r_{13})$ is the 'reduced' 3-point correlation 
function or simply known as 3-point correlation function. The terms in 
eq.~\ref{thirdeq} depending on the 2-point correlation function reflect the excess
number of triplets one has when there is an excess of pairs over a random 
distribution. Then the term $ \zeta(r_{12},r_{23},r_{13})$ is simply the excess
of triplets over a distribution with a given 2-point correlation function. When
approximating by a continuous distribution of objects we have,
\begin{equation}
\zeta(r_{12},r_{23},r_{13})=<\delta (\vec r_1)\delta (\vec r_2)\delta (\vec r_3)>.
\label{fourtheq}
\end{equation}
\noindent 
Notice that we have just written $\mbox{$<\delta (\vec r_1)\delta (\vec r_2)\delta (\vec r_3)>$}$ instead of $<\delta (\vec r_1)\delta (\vec r_2)\delta (\vec r_3)>_c$, the cumulant or reduced moment, but because of the definition of 
$\delta (\vec r)$ this one has a zero mean so both types of moments coincide.
One interesting characteristic of the relation between moments and cumulants is
the fact that for the case of a Gaussian distribution all the cumulants of 
order larger than $2$ are zero [see, for instance, ~\cite{SO94,BCGS01}].


In the study of the 3-point correlation function we will be interested in
 computing the $Q_3$ parameter in real space in N-body
simulations. The $Q_3$ parameter, introduced in ~\cite{GP77},  is defined as 
follows,
\begin{eqnarray}
Q_3 &=& \frac{\zeta(r_{12},r_{23},r_{13})}{\zeta_H(r_{12},r_{23},r_{13})} \\
\zeta_H  &\equiv& 
{\xi (r_{12})\xi(r_{23})+\xi(\vec r_{12})\xi(r_{13})+\xi(r_{23})\xi(r_{13})}.
\label{fiftheq}
\end{eqnarray}
\noindent
where we have introduced a definition for the "hierarchical" 
3-point function $\zeta_H$.
The $Q_3$ parameter was thought to be a 
good description of observations by being 
constant (see ~\cite{PB80}), a result that is usually referred to as the
 hierarchical  universe. As we will see below, in the weakly non-linear
regime (or large scales compared to the correlation length), the
universe is not exactly hierarchical as $Q_3$ is not quite constant,
but this is a good approximation in some cases.

\subsection{Tree-level contribution.}

\begin{figure}
\centering{
{\epsfxsize=9cm \epsfbox{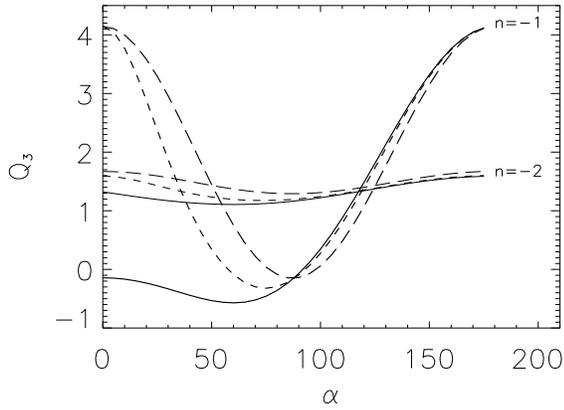}}}
\caption[junk]{Plot of the value of $Q_3(r_{12},r_{23},\mu)$ as a 
function $\alpha$ (recall that $\mu=\cos\alpha$), 
the angle between $\vec r_{12}$ and 
$\vec r_{23}$, as obtained in perturbation theory (up to first non-null terms) 
for scale free models. Solid lines are for values 
of $u= r_{23}/r_{12}=1$, short-dashed ones are for $u=2$ and long-dashed 
correspond to $u=10$. Different lines' thicknesses correspond to different 
spectral indexes $n$ as shown.}
\label{fig2}
\end{figure}

In perturbation theory (starting from Gaussian initial conditions),
the first non-null 
contribution to the 3-point function is well-known 
and it can be even expressed in analytical form in the case of scale 
free (power law) power spectrum: $P(k)=Ak^n$ [see 
~\cite{Peebles80,Fry84,BCGS01} and references therein],
\begin{eqnarray} 
\lefteqn{\zeta(r_{12},r_{23},\mu)= \left[ \frac{10}{7}+\frac{n+3}{n}\mu 
(\frac{r_{12}}{r_{23}}+\frac{r_{23}}{r_{12}})+ \right.}  \\ 
\lefteqn{\left.\frac{4}{7}\frac{(3-2(n+3)+(n+3)^2\mu ^2)}{n^2} \right]
\xi(r_{12})\xi(r_{23})+\mbox{permutations}\nonumber}
\label{sixtheq}
\end{eqnarray}
$\zeta(r_{12},r_{23},\mu)=\zeta(r_{12},r_{23},r_{13}(\mu))$ where $\mu$ 
is the cosinus of the angle between $\vec r_{12}$ and $\vec r_{23}$, say 
$\alpha $. An
 interesting property of the power law case  is that
 $\zeta(r_{12},r_{23},r_{13})$ (and also $Q_3$) only 
depends on the ratios of distances $r_{12}$, $r_{23}$, $r_{13}$. As these 
distances bear the additional constraint of forming a triangle the dependence of 
$\zeta(r_{12},r_{23},\mu)$ is reduced to only $2$ parameters, for example,
$u\equiv r_{23}/r_{12}$ and $v\equiv (r_{13}-r_{23})/r_{12}$ used by Peebles
and collaborators, Jing and B\"orner and Buchalter and Kamionkowski or the 
parameters we will use here, 
$u\equiv r_{23}/r_{12}$ and $\mu=\cos \alpha$ where  $\alpha$ is the angle 
defined above. We have chosen these latter 
parameters and not the former ones  
because they are in more direct relation to the way in which the lists employed by
 our algorithm (explained in subsection \ref{thealgo}) are built. In figure  ~\ref{fig2} we show how 
$Q_3(r_{12},r_{23},\mu)$ 
changes as a function of $\alpha$ for some values
 of $u=r_{23}/r_{12}=1,2,10$. The thicker lines correspond to a model with 
spectral index $n=-1$ and the 
thinner ones to a model with $n=-2$. It is noticeable how the $Q_3$ flattens 
with the decrease of the spectral index $n$. For each model parametrised by the
spectral index there are three lines corresponding to different $u$ values. The
solid ones are for $u=1$ (isosceles triangles), the short-dashed ones are for
$u=2$ and the long-dashed ones are for $u=10$.
  
\subsubsection{CDM models}

In the case of CDM power spectra the solution is not analytical but can be
 expressed in terms of some basic functions that can be easily computed numerically,
\begin{eqnarray} 
\lefteqn{\zeta(r_{12},r_{23},\mu)= \frac{10}{7}\xi(r_{12})\xi(r_{23})+ 
\nonumber }\\
\lefteqn{+\frac{4}{7}\big(-3\frac{\phi'(r_{12})\phi'(r_{23})}{r_{12}r_{23}}
-\frac{\xi(r_{12})\phi'(r_{23})}{r_{23}}-\frac{\xi(r_{23})\phi'(r_{12})}{r_{12}}+\nonumber }\\
\lefteqn{+\mu ^2(\xi(r_{12})+
3\frac{\phi'(r_{12})}{r_{12}})(\xi(r_{23})+3\frac{\phi'(r_{23})}{r_{13}})\big)-\nonumber} \\
\lefteqn{-\mu(\xi'(r_{12})\phi'(r_{23})+\xi'(r_{23})\phi'(r_{12}))+\mbox{permutations},}
\label{seventheq} 
\end{eqnarray}
\noindent
where 
\beq \phi(r)=\int d\vec k \frac{P(k)}{k^2}\frac{\sin(kr)}{kr} 
\eeq 
and $P(k)=A ~k~T(k)^2$ 
is the linear power spectrum for CDM models with primordial
spectral index of the Harrison-Zeldovich type $k$, and $A$
is a normalization factor. We 
take a transfer function $T(k)$ of
the Bond \& Efstathiou type:
\beq
T(k)=\Big(1+\big(ak+(bk)^{1.5}+(ck)^2\big)^{\nu}\Big)^{-1/\nu}
\eeq
with $a=6.4/\Gamma\mpc$, $b=3/\Gamma\mpc$, $c=1.7/\Gamma\mpc$, $\nu=1.13$.
$\Gamma\approx \Omega _m h$ is the so-called 'shape parameter' and $h$ is
Hubble's constant in $100 Kms^{-1}/Mpc$ units). This above Eq.[\ref{seventheq}]
is also identical to
results in ~\cite{JB97,GB98} once a relation between their functions and ours
is established. 

\begin{figure}
\epsfxsize\hsize\epsffile{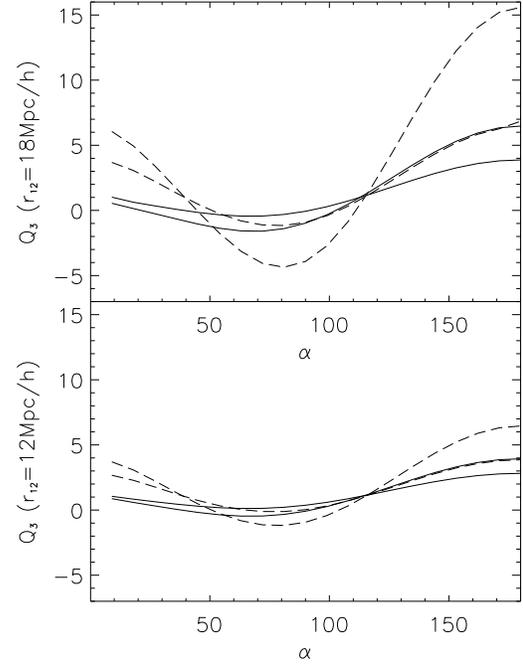}
\caption[junk]{Plot of the value of $Q_3$ as a 
function of $\alpha$, as obtained in perturbation theory (up to first non-null 
terms) for $CDM$ models with shape parameter $\Gamma=0.2 $, thick lines, and
$\Gamma=0.3 $, thin lines. Solid lines are for values 
of $u= r_{23}/r_{12}=1$ and long-dashed correspond to $u=2$.}
\label{fig3}
\end{figure}

It is interesting to note that for $CDM$ power spectra
$\zeta(r_{12},r_{23},\mu)$ is no longer a function of the ratios of $r_{12}$,
$r_{23}$ and $r_{13}$, as it was in the case of scale free power spectra, and
so $3$ variables are now needed (the extra variable defines an "effective"
index in the analogy to the power-law case).  In figure ~\ref{fig3} we show
this dependency on $r_{12}$ by displaying two plots with the same range in the
ordinate axis. On the top (bottom) panel we show the computed values for $Q_3$
in perturbation theory in the case $r_{12}=18\mpc$ ($r_{12}=12\mpc$). The
solid lines correspond to isosceles triangles ($u=1$) whereas long-dashed
lines correspond to $u=2$. The thick lines represent cosmological models with
a shape parameter $\Gamma=0.2$ and the thinner lines represent cosmological
models with $\Gamma=0.3$. Notice how for small scales (the lower plot) the
curves become flatter. It is also quite noticeable from figure ~\ref{fig3} how
increasing the shape parameter, $\Gamma$, mimics the effects of doubling the
value of $u$ for angles larger than, say, $70^0$ but the difference is clear
for smaller angles. In summary, both increasing the shape parameter and
increasing scales make curves have more shape but the effects due to one case
or the other can, in principle, be disentangled by looking at different ranges
of angles above and below approximately $70^0$.

Note the crossing of lines at $\alpha \simeq 120^0$ in both power-law and CDM
cases, even for different values of $u$. The crossing occurs at $\alpha \simeq
116$ and $Q_3 \simeq 1.18$, which seems to be a characteristic feature of the
tree-level gravitational evolution.

As we mentioned above this is the first non-null contribution (also called
'tree-level') to the 3-point correlation function but there is an interesting
feature in considering the next non-null terms, the so-called '1-loop' terms.

\subsubsection{Up to 1-loop contributions.}

Consider the second non-null terms in both
$\zeta(r_{12},r_{23},\mu)$ and the denominator of $Q_3(r_{12},r_{23},\mu)$ 
(the products of 2-point correlation functions), which were 
called $\zeta_H(r_{12},r_{23},\mu)$ in eq.\ref{fiftheq}, 
the hierarchical 3-point correlation function, we have schematically,
\begin{equation}
Q_3=\frac{\zeta^{(0)}+\zeta^{(1)}}{\zeta_H^{(0)} +\zeta_H^{(1)}}
\label{eightheq}
\end{equation}
\noindent
where $\zeta_H^{(0)}$ is the first non-null term in perturbation theory of the
denominator of $Q_3$ and $\zeta_H^{(1)}$ the next non-null term.  For the
explicit content of these terms in Fourier space see ~\cite{S97}. For example,
formulae from eq.[32] to eq.[36] in ~\cite{S97} give the expression for the
$\zeta^{(1)}$ term in Fourier space (his $B^{(1)}(k_1,k_2,\tau)$) and formulae
eq.[39] and eq.[24-26] give the expression for the Fourier space counterpart
of $\zeta_H^{(1)}$ (his $\Sigma^{(1)}(k_1,k_2,\tau)$). $\zeta^{(0)}$ and
 $\zeta_H^{(0)}$ terms depend on the amplitude
of the power spectrum squared, $A^2$, whereas $\zeta^{(1)}$ and
$\zeta_H^{(1)}$ depend on $A^3$ (explicitly in ~\cite{S97}). We will make use
 of this property in order to
compare these terms extracted from simulations to the first non-null terms in
perturbation theory
($Q_3^{(0)}(r_{12},r_{23},\mu)=\zeta^{(0)}/\zeta_H^{(0)}$).

\section{N-body results}

In this section we first describe briefly the N-body simulations that
we will use. We next present two of the
new contributions of this paper.  The first of these contributions
 consists on an algorithm that  drastically reduces the amount of
 time necessary to compute the correlation functions in N-bodies simulations 
with millions of particles. The second main idea refers to devising a way to 
disentangle non-linearities in the simulations by combining different
output times.

\subsection{Description of the simulations}
\label{Des}

The N-body simulations we will use here are evolved 
using the $P^3M$ code of ~\cite{E85}. We have considered
the following cases, 
A flat $\Lambda$CDM
model with $\Omega_m=0.2$ and $\Gamma=0.2$ in a cube of side $378\mpc$ 
containing $126^3$ particles. 
We have also considered a $SCDM$ ($\Omega_m=1$, $\Gamma=0.5$) model in a cube 
of side $378\mpc$ also containing $126^3$ particles. We have
$5$ realisations for each model and consider two different
output times $\sigma _8=0.5$ and $\sigma _8=1$. 
Table ~\ref{tab1} summarizes the main parameters
in the simulations [see \cite{BGE95} for more details on these runs].

We have also use a new set of simulations with CDM shape $\Gamma=0.25$ but
with $\Omega_{m}=1$ (called $hCDM$ in table ~\ref{tab1}). These new
simulations can be used to test the sensitivity to $\Omega_m$ with independent
of the shape of $P(k)$. They have a box size of $300 \mpc$ and twice the
particle resolution of the previous sets ($N=200^3$ particles). Thus they can
also be used to check for shot-noise and resolution effects. The computation
of gravitational dynamics was started at $z=20$ (instead of $z=10$ in the
other CDM sets), so that they are less sensitive to possible Zeldovich
Approximation transients (see Baugh, Gazta\~naga \& Efstathiou 1995,
Scoccimarro 1998).  There are 5 realisations of the $\sigma_8=0.5$ output.

\begin{table}
\begin{tabular}{|c||c|c||c|} \hline
                 & $\Lambda CDM$  & $SCDM$ &  $hSCDM$\\ \hline
\hline
$\Omega _m$         &  0.2     &  1      & 1  \\ \hline
$\Omega _{\Lambda}$ &  0.8     &  0      & 0  \\ \hline
$h$                &  0.7     &  0.5    & 0.3 \\ \hline
$\Gamma $          &  0.2     &  0.5    & 0.25 \\ \hline
$Npar$             &  $126^3$ & $126^3$ & $200^3$ \\ \hline
$Lsize$            & $378\mpc$&$378\mpc$&$300\mpc$ \\ 

\hline
\end{tabular}
\caption[junk]{Simulations' main parameters.}
\label{tab1}
\end{table}

\subsection{The algorithm to estimate correlations.}
\label{thealgo}

Once simulations have  evolved gravitationally from initial conditions  we 
compute the 2-point and 3-point correlation functions. To do this
with $N$ particles, we would  naively  
require $N^2$  and $N^3$ operations for the 2-point 
and 3-point function respectively.  
The case of $N=200^3$ gives  $5 \cdot 10^{20}$ operations while the case of
$N=126^3$ gives $\approx 8 \cdot 10^{18}$ operations for the 3-point 
function. These numbers are out of reach for an 
ordinary computer if we bear in mind that many simulations 
(different realisations multiplied by output times) are needed. 

Based on ~\cite{FG99} we devise a method which dramatically reduces the number
of operations. The first step is to discretize the simulation box
into $Lsize^3$ cubic cells. We assign each particle to a node of this new
latticed box using the Nearest Grid Point particle assignment (NGP).  
We thus obtain a density field over the $Lsize^3$ lattice, 
resulting from just counting particles in each node.

To compute now the 2-point and 3-point correlation functions in the lattice
would naively take $(Lsize^3)^2$ and $(Lsize^3)^3$ operations respectively.
This means a considerable reduction of operations when $Lsize<N^{1/3}$ or,
equivalently, when the mean cell density ($N/Lsize^3$) is larger than $1$.
As we have seen in section ~\ref{the2} both the 2-point and 3-point
correlation functions depend only on lengths. This fact is used in the next
subsection to devise a method which diminishes the number of operations even
further.

\subsubsection{The 2-point correlation function.}

The 2-point correlation function depends only on the distance between two
points (labeled '1' and '2'). In general, we want to consider a finite set of
fixed distances which are much smaller than the box size (as otherwise our
statistics will be dominated by sampling effects). So we would like to learn
how to count only pairs of points separated by a fixed distance. It is then
convenient to define a list of neighbors to each point.  This list is formed
by members of the lattice which are inside a spherical shell of a certain
width of radius say $r_{12}$. In the lattice, this list is given by a single
list of "displacement vectors" for all nodes.  In a more general situation
(where particles are not placed on a lattice) this list is different for each
particle. In both cases the lists are pre-computed. In the case of the
lattice, to calculate the list of "displacement vectors" takes a negligible
amount of computing time.

In order to count only these pairs, we run through all nodes in the cubic
latticed box and for each node we loop over the attached list of neighbours.
The saving in computing time resides in the fact that for each node we do not
compute the $Lsize$ other nodes but only those nodes which belong to the list
(say $m_{12}$). The number of operations is reduced from $(Lsize^3)^2$ to
$(Lsize^3) \cdot m_{12}$, where $m_{12}$ (the number of the nodes belonging to
a spherical shell) is always much smaller than $Lsize ^3$ (the nodes in the
box; the diameter of the shell is restricted to be smaller than the side of
the box). To show some numbers, let us consider the case with $Lsize=63$.
Naively the number of operations would be $(63^3)^2\approx 6.2\cdot 10^{10}$.
When we are interested in a fixed scale, say, for example $r_{12}=3$ (in
lattice units) there are 66 entries (i.e. nodes) in the list of
"displacement vectors", so the number of operations would be $(63^3)\cdot
66\approx 1.6\cdot 10^7$. An additional saving of a factor of two can be
achieved by realising that the ordering of the pairs is not relevant. This
means in practice that we only need a semi-sphere of neighbours in our list.

One might think that this saving is ficticious as it would take the
same time to loop over all particle pairs as to loop over 
all possible pair distances using the neighbours' list. But recall
that in practice the scales of interest are always $r_{12} << Lsize$,
so that it does not make sense to loop over all pair distances

The advantage of this method in computing time  becomes
far more evident in the case for the 3-point and higher
order correlation functions.

\subsubsection{The 3-point correlation function.}

\begin{figure}
\epsfxsize\hsize\epsffile{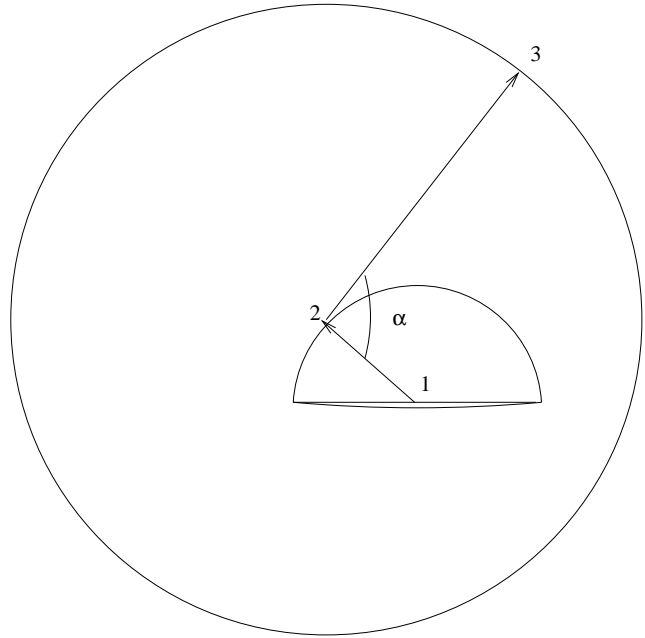}
\caption[junk]{Schematic plot for the method explained in the text to compute
the 3-point correlation function in simulations. From point labeled '1' we
trace the shell of a semi-sphere of a finite thickness with radius $r_{12}$. 
From each point of the shell (points 
labeled '2') we trace another spherical shell with
radius $r_{23}$ (a 
whole sphere this time).}
\label{fig1}     
\end{figure}

The 3-point function depends only on the length of the three vectors
formed by the 3 point-particles, that is on the sides of a triangle formed
by these 3 point-particles being at the vertexes. We can also express this
dependency as a dependency on any two vectors, chosen from the three
available, and the angle between them $\alpha$. Then the method now consists
of using the same lists of neighbours, as for the 2-point function's case, 
with two
nested spheres, each one of radius equal to the length of one of these two
vectors.  Thus, we run through every single lattice node in our simulations
boxes (node labeled '1' in figure ~\ref{fig1}). To each node in the
lattice we run over the list of neighbours within 
a semi-sphere shell of radius $r_{12}$ and a certain width (all these nodes
are candidates for being labeled as '2' in figure ~\ref{fig1}). Each of the
new '2' nodes of the semi-sphere are now centers for another spherical shell of
radius $r_{23}$ and a certain width (these new nodes are labeled '3' in figure
~\ref{fig1}). In summary, the list of triplets separated by distances 
$r_{12}$ and $r_{23}$ is built from two nested
list of  neighbours of separations $r_{12}$ and $r_{23}$
(see figure ~\ref{fig1}). As in the case of the 2-point function,
the first semi-sphere (as opposed to a full sphere) is used to avoid
double counting pairs.

This method of building the neighbours list associated to a node in the box
makes us save even more time than in the case of the 2-point correlation
function. The number of operations to be done in the case for fixed scales are
reduced to $(Lsize^3)\cdot m_{12}/2\cdot m_{23}$ instead of $(Lsize^3)^3$,
where $m_{12}/2\cdot m_{23}$ is the number of nodes in the neighbours list. To
write down some numbers let us take the example of latticing the box with
$Lsize=63$. To compute the 3-point correlation function naively would take
$(63^3)^3\approx 1.6\cdot 10^{16}$ operations in the lattice (recall that this
was as large as $5 \cdot 10^{20}$ from the raw particle positions) whereas by
using our method of associating a neighbours list for fixed scales, say for
instance $r_{12}=r_{23}=3$ (in lattice units), would take $(63^3)\cdot 33\cdot
66\approx 5.4\cdot 10^8$ operations.

This method can be trivially extended to higher order correlations.

\subsubsection{Generalities of the algorithm.}

One characteristic of our algorithm is that shells, from which we build our
lists, are distorted. This is always the case when trying to obtain
spherical objects in a cubicly gridded box.  Obviously, the distortion in the
spherical nature of shells is going to diminish with the increase of the
radius in terms of how many cells long the latter is. We have studied these
effects by using different size cells (or lattices) to study  the
same physical side.  For example, in the case of
$\Lambda CDM$ model with $\Omega _m=0.2$ simulation, which has a box of
$378\mpc$, we have used a lattice of $Lsize=63$ nodes with a radii
$r_{12}=3$ and $r_{23}=3$ in cells units (to study the physical scales
$r_{12}=r_{23}=18\mpc$). We also latticed the simulation with with $Lsize=126$
and a radii $r_{12}=r_{23}=6$ (in cell units), which
corresponds to the same physical distance.
We find that the differences are negligible when we use
 3 or more cell units for the distance. To reduce the number of operations we
want $Lsize$ to be as small as possible, we therefore
choose $Lsize$ so that the
scale of interest corresponds to $r=3$ cell units. The thickness
of the shell is also a parameter which affects the distortion, 
as some cubic cells
may be missed out if we choose this parameter to be too small and too many
cells will be included if this parameter is too large. We have studied these
effects and have adopted a compromise of  a 
thickness of $20\%$ in cell side units.

\subsection{Disentangling nonlinearities in the 
2-point and 3-point correlation function.}
\label{disen}

As can be seen from its definition $Q_3$ depends on two sides of the triangle
formed by the three points and on the angle between them. Examining the
$\alpha$ dependence it is clear that 
nonlinearities are going to be important when $\alpha \rightarrow 0$.
When the third side approaches zero the 2-point correlation function on this
scale is going highly non-linear. On the other extreme, at large scales, it is
also possible that $Q_3$ diverges because the denominator approaches zero.
This is unavoidable in CDM models, as the 2-point correlation becomes negative
at some large scale which is not very far from the non-linear scale.
This divergence of $Q_3$ has promoted proposals for
'better behaved' normalizations for $\zeta$, but with similar scaling as
$Q_3$. This is the case of ~\cite{BK99} where they define $Q_v\equiv
\zeta(r_{12},r_{23},\mu)/\sigma_8^2$ with:
\begin{equation}
\sigma_8^2\equiv \bar\xi_2(r_{12}=0, R_f=8) \equiv 
\frac{1}{(2\pi)^3}\int d\vec k P(k)W(kR)^2
\label{sigma8}
\end{equation}
and
 $W(kR)$ is the Fourier transform of the window function.
Non-linearities affects both the numerator and denominator of $Q_3$. We
need to separate the different non-linear components in $Q_3$ if we want
a consistent comparison of PT with  simulations.
 In order to estimate the effects of mild nonlinearities
in the simulations we devise the following method, based on the
scaling of both the 3-point and 2-point correlation functions with the
amplitude of the linear power spectrum (or growth factor) in perturbation
theory. 

As explained in the previous section the tree-level (that is, the first
non-zero term) contribution to the 3-point function depends on the square of
the amplitude $A^2$ of the power spectrum and the one-loop level
(next order in perturbation theory) depends on $A^3$. We can do exactly the
same in the denominator developing it up to order $A^3$. 
Imagine now that we get two
outputs of our simulations with different amplitudes, i.e. different
$\sigma_8$. For example,
 one with $\sigma_8=0.5$ and another one with
$\sigma_8=1$. This is a difference in amplitude of a factor of 4. 
We can then pose a very simple system of two equations with two unknowns,

\begin{eqnarray} 
\lefteqn{x+y=n_1\nonumber }\\
\lefteqn{\frac{x}{16}+\frac{y}{64}=n_2}
\label{nintheq}
\end{eqnarray}
\noindent
where $x$ stands for $\zeta^{(0)}$ or $\zeta_H^{(0)}$ (i.e., tree-level terms),
 $y$ stands for $\zeta^{(1)}$ or $\zeta_H^{(1)}$ (i.e., 1-loop terms)
 and $n_1$ and $n_2$ are the results we obtain for the 3-point function with
simulation output $\sigma_8=1$ and $\sigma_8=0.5$ respectively.
 So, in
this way we are able to obtain the tree-level ('x') both in the numerator and
denominator. 

This could obviously be extended to higher order corrections if we had more
output times. Note that these set of equations will always return a solution.
The solution will only be meaningful if deviations from the tree-level are
small enough at the order of perturbations we are considering. One way to test
the consistency of the solution is to verify that we are able to recover the
tree-level solution ('x') from the different outputs. Clearly, on small,
strongly non-linear scales, the perturbative approach will break down and the
solution will be meaningless (in such case we do not expect to recover the
tree-level solution for 'x').

\section{Comparison to simulations}
In this section we compare the results to actual simulation
 values. We first analyze different triangles configurations (isosceles 
and equilateral) using our algorithm (explained in subsection ~\ref{thealgo}) 
and examine how they compare to theoretical values in 
subsections ~\ref{shap} and ~\ref{equi}. In subsection ~\ref{loop} we apply our
method to disentangle possible 'non-linearities' due to 1-loop terms and obtain
a the tree-level from simulations (as described in ~\ref{disen}).


\subsection{Shape dependence for isosceles}
\label{shap}

\begin{figure}
\centering{
{\epsfxsize=9cm \epsfbox{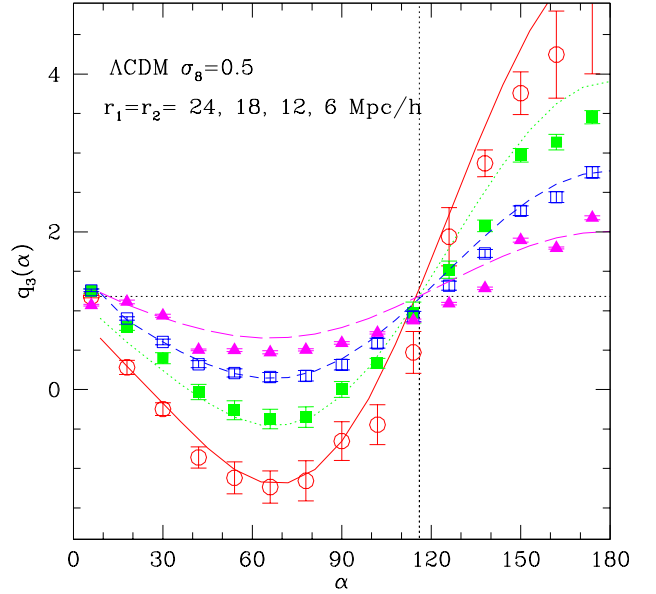}}}
\caption[junk]{Plot of the value of $Q_3(r_{12},r_{23},\alpha)$ as a 
function $\alpha$ in $\Lambda CDM$ models with $\sigma_8=0.5$, 
the angle between $\vec r_{12}$ and 
$\vec r_{23}$, for different values of 
$|\vec r_{12}|=|\vec r_{23}|=24,18,12,6\mpc$ 
(opened circles, closed squares, open squares and
triangles). The corresponding
lines (continuous, dotted, short-dashed and long-dashed) show the
tree-level perturbation theory.}
\label{q3c80dat3}
\end{figure}

\begin{figure}
\centering{
{\epsfxsize=9cm \epsfbox{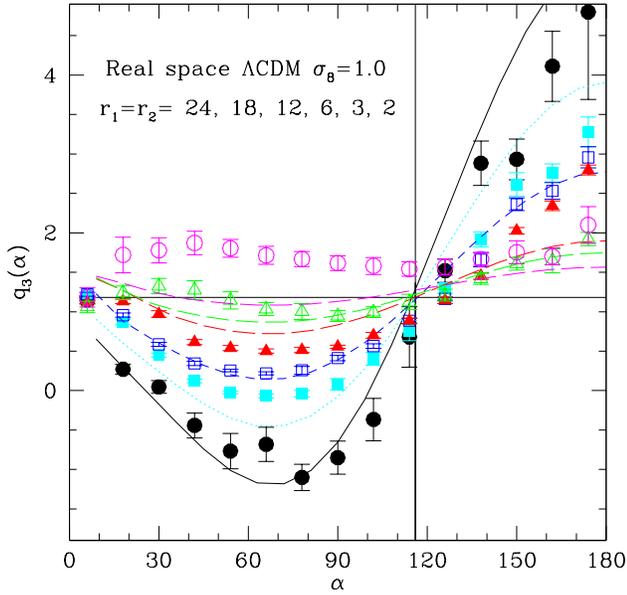}}}
\caption[junk]{Same as fig.\ref{q3c80dat3} for the $\sigma_8=1$ output.}
\label{q3c80dat6}
\end{figure}

Figures \ref{q3c80dat3} and \ref{q3c80dat6} show the estimations of $Q_3$ in
the $\Lambda CDM$ model for isosceles triangles $| \vec r_{12}|=|\vec
r_{23}|=24,18,12,6\mpc$ at two different output times: $\sigma_8=0.5$ in
fig. \ref{q3c80dat3} and $\sigma_8=1.0$ in fig. \ref{q3c80dat6}. The figures
with error-bars show the mean and variance in 5 simulations, while the different
lines show the corresponding
leading order (tree-level) PT prediction. As can be seen in
fig.\ref{q3c80dat6} the tree-level is recovered well for all scales $|\vec
r_{12}|=|\vec r_{23}|> 6\mpc/h$ in the $\sigma_8=0.5$ output.  At $|\vec
r_{12}|=|\vec r_{23}|=6\mpc/h$ there are significant deviations from the
leading order results indicating that loop corrections are important.

In the $\sigma_8=1.0$ output (fig.\ref{q3c80dat6}) the discrepancies are
significant even for $|\vec r_{12}|=|\vec r_{23}|= 24\mpc $. Note how
deviations from tree-level tend to smooth or wash out the shape dependence for
$|\vec r_{12}|=|\vec r_{23}|> 12\mpc$, while on smaller scales there is more
shape dependence than in the tree-level, in the sense that $Q_3$ is smaller at
smaller $\alpha$ and larger at larger $\alpha$.  This tendency is also true
when comparing the shape of two outputs at a given scale: i.e. the later
outputs have more shape dependence than the earlier outputs at small scales
($< 12 \mpc$), while the opposite tendency occurs for the larger scales.  This
effect looks similar to the "previrialization effect" present in the 2-point
function ~\cite{SF96,LJBH96}. A similar trend was also found in the bispectrum
~\cite{S97}.  The idea is that there is a change of sign in the 1-loop
contribution which, for a power-law initial spectrum, is related to a critical
value in the spectral index.  We will study this in more detail below.

\subsection{Equilateral configurations}
\label{equi}
\begin{figure}
\centering{
{\epsfxsize=9cm \epsfbox{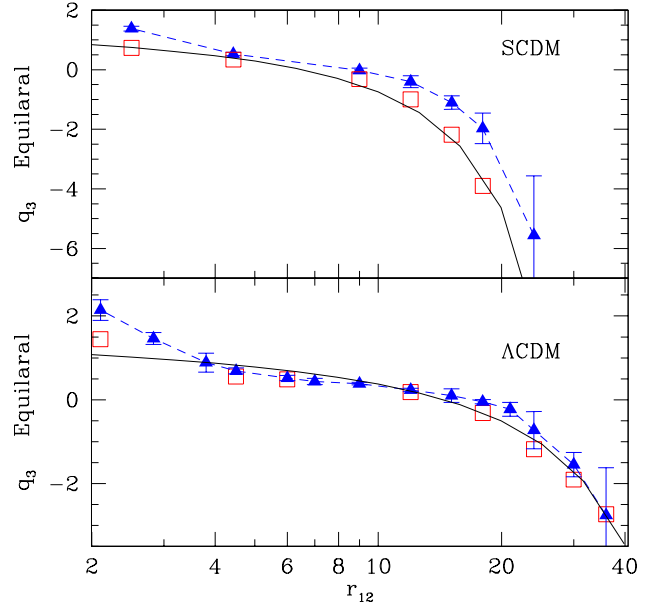}}}
\caption[junk]{The normalized 3-point funciton $q_3$
 for equilateral triangles as a funciton of the
triangle side. The continuous line correspond to the
PT tree-level predictions 
while the triangles  and errors show the
corresponding estimation for the mean and variance of 5 realisations
in the  $\sigma_8=1.0$ output. Opened squares show the mean of the
5 realisations in the  $\sigma_8=0.5$ output.
The top panel correspond to
the $SCDM$ model while the bottom panel
shows the $\Lambda CDM$ results.}
\label{q3eqall}
\end{figure}

Fig.\ref{q3eqall} shows $Q_3$ for equilateral triangles (i.e. fixed $|\vec
r_{12}|=|\vec r_{23}|$ and $\alpha=60^0$) as a function of the triangle side
($|\vec r_{12}|=|\vec r_{23}|$), for both the $\sigma_8=1.0$ output
(triangles) and the $\sigma_8=0.5$ (squares), in both the $SCDM$ (top panel)
and the $\Lambda CDM$ (bottom panel) model. The dashed line just connects the
points at different scales. Also shown as a continuous line is the leading
order (tree-level) contribution.

For both models, the $\sigma_8=0.5$ output (squares) shows good agreement with
the PT results.  For $\sigma_8=1.0$ the $SCDM$ shows significant deviations
from PT at all scales. On the largest scales the 2-point function goes to zero
at smaller scales in the $SCDM$ model which explains why the errors are much
larger around $20 \mpc$ in this model.

The $\Lambda CDM$ model shows larger $Q_3$ values than PT at $r_{12} \simeq 20
\mpc$ and smaller values at $ r_{12} \simeq 6 \mpc$. This is related to the
change in shape away from the leading order result noted in the subsection
above
(e.g. see fig.\ref{q3c80dat6}).  More in detail, at large weakly non-linear
scales, $ r_{12} > 12\mpc$ where the 2-point function is steeper (with a slope
$\gamma<-2.2$) PT corrections on $Q_3$ are above tree-level, while for $
r_{12} <12 \mpc$ the 2-point function is flatter (with a slope $\gamma>-2.2$)
PT corrections on $Q_3$ are below tree-level.  On smaller scales, i.e.
$r_{12}<4\mpc$, non-linear effects take over and $Q_3$ increases again above
the PT results.
These deviations from the leading order results are small, but
significant given the sampling errors.

\subsection{Loop corrections}
\label{loop}

\begin{figure*}
\epsfxsize\hsize\epsffile{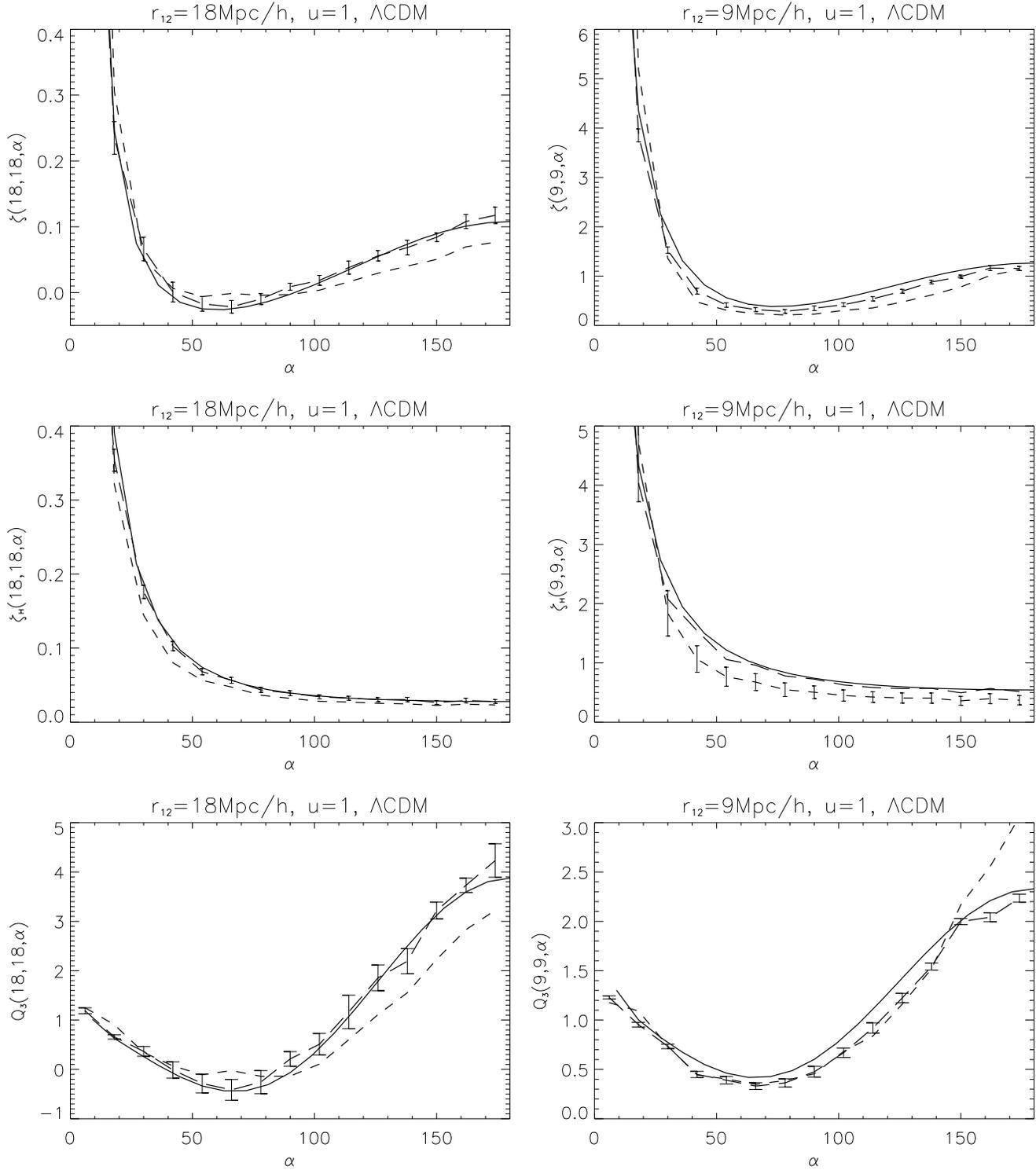}
\caption[junk]{Recovering the tree-level and 1-loop contributions to the 
numerator (top) and denominator (middle) of $Q_3$, and also for $Q_3$ itself
(bottom). The cosmological model
is flat $\Lambda CDM$ with $\Omega_m=0.2$. The left (right) column correspond 
to isosceles triangles with equal side being $18\mpc$ ($9\mpc$) long.
The solid lines represent tree-level 
PT prediction, the long-dashed lines shows tree-level recovered for the
simulations while the short-dashed corresponds to the estimated
1-loop level in the simulations (at $\sigma_8=1$). For clarity, 
only the larger error-bars in either of the two dashed lines are shown.}
\label{fig4} 
\end{figure*}

In figure ~\ref{fig4} we show our method to recover the tree-level and the
first-loop contributions for the denominator and numerator of $Q_3$, i.e.
for  $\zeta $,
and $\zeta_H $ respectively, as explained in subsection ~\ref{disen}. All the
plots in figure ~\ref{fig4} are for a flat $\Lambda CDM$ model 
with $\Omega_m=0.2$.
 The plots on the left column are for isosceles triangles
with equal side of $18\mpc$ long whereas the ones on the right correspond
to isosceles triangles with equal side of $9\mpc$ long. From top
to bottom, we show  $\zeta $ (the 3-point function), 
$\zeta_H $ (the hierarchical combination in eq.[\ref{fiftheq}]) 
and $Q_3$ (the ratio of both). 
In all the plots of figure ~\ref{fig4} the solid lines correspond to
the tree-level theoretical predictions from PT, the
long-dashed and short-dashed lines 
correspond to the tree-level and 1-loop ($\sigma_8=1$) estimation from
the 2 simulation output times, as described in eq.~\ref{nintheq}.

We can now address the question raised in the previous
subsections of whether or not the relative change of sign in the deviations
of $Q_3$ from tree-level PT (e.g. shown in fig.\ref{q3c80dat6} and
fig.\ref{q3eqall})  is related to the "previlarization effect" on the 2-point
function.
 
In the middle row plots we can see 
how for both scales ($18\mpc$ and $9\mpc$) the 1-loop contribution to $\zeta_H$
(short-dashed line) is lower than the tree-level one (long-dashed line); this
can be interpreted in both cases as the well-known effect of 
previrialization found in the 2-point function ~\cite{SF96,LJBH96}.
But here, we have to bear in mind that $\zeta_H$
 is a product of two 2-point correlation functions.

 In the top panels one can see how the 1-loop contribution to $\zeta$
 (short-dashed lines), are larger than the tree-level contribution around
 $\alpha \simeq 60$ deg.  for $18\mpc$, while they are smaller than tree-level
 for $9\mpc$. This effect is the one mostly responsible for the relative change
 shown in $Q_3$ (smoothing/enhancing the shape with respect to the
 tree-level).  Thus, we see that this tendency is not just given by the same
 critical scale index in the 2-point previrialization, as here we found a case
 where there is change of sign at the 3-point function level (when comparing
 $18\mpc$ with $9\mpc$ results), but not for the 2-point function. This is not
 surprising, as there are different PT terms contributing to each of these
 expressions.

\begin{figure*}
\epsfxsize\hsize\epsffile{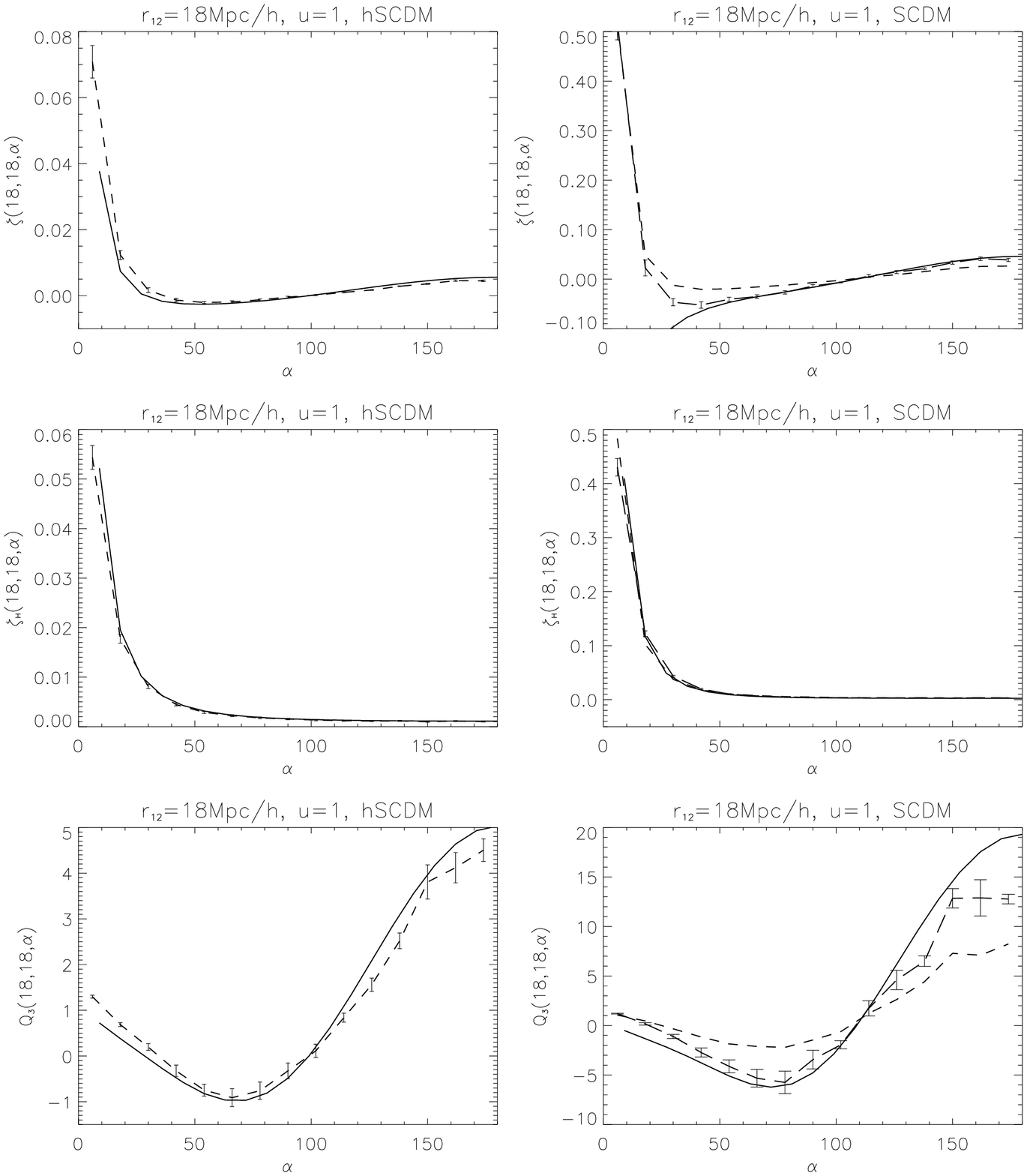}
\caption[junk]{On the left column we show results for $\zeta$, $\zeta_H$
and $Q_3$ in the $hSCDM$ model
with  $\Omega_m=1$ $h=0.3$, corresponding to isosceles 
triangles configurations with equal side $18\mpc$.
The dashed line with error-bars correspond
to the $\sigma_8=0.5$ output, the continuous line shows the
leading order PT prediction.
 On the right column we show the same case for the model
 $SCDM$ ($\Omega_m=1$,$h=0.5$ ). Here
the criteria for lines are identical to those for
figure ~\ref{fig4}.}
\label{fig5} 
\end{figure*}

In figure ~\ref{fig5} we compare the 3-point correlation function ($\zeta$),
the hierarchical 3-point correlation function ($\zeta _H$) and the full $Q_3$
parameter for the model $hSCDM$, on the left column, which has $\Omega_m=1$,
$h=0.3$ (see table ~\ref{tab1}) to the corresponding values
in the $SCDM$ model, on the right column, with $\Omega_m=1$, $h=0.5$
(the old standard, see table ~\ref{tab1} ).  
 As before the solid lines correspond to the
theoretical results in PT.  
The short-dashed shows the  $\sigma_8=0.5$ output. On the left panel,
the short-dashed and long-dashed 
lines correspond to the tree-level and 1-loop reconstruction in the
SCDM case. The purpose of this figure is to illustrate how the $\zeta$,
$\zeta _H$ and $Q_3$ functions change with the cosmological model. The $hSCDM$
model has an approximate shape parameter for the power spectrum $\Gamma=0.25$
whereas the $SCDM$ model has a $\Gamma=0.5$. Notice
that although the shape of the curves of the left and right panels
in figure ~\ref{fig5} are quite similar the actual values in the range of the
plots are  different.

Figure ~\ref{fig5} also illustrates 
how 1-loop corrections (shown as short-dashed lines) in the SCDM
tend to flatten the
$Q_3$ curve (see bottom right plot), in a similar, but more
dramatic, way as we found in the $\Lambda CDM$ model
in figure ~\ref{fig4}. 

\subsection{Other configurations}

\begin{figure*}
\epsfxsize\hsize\epsffile{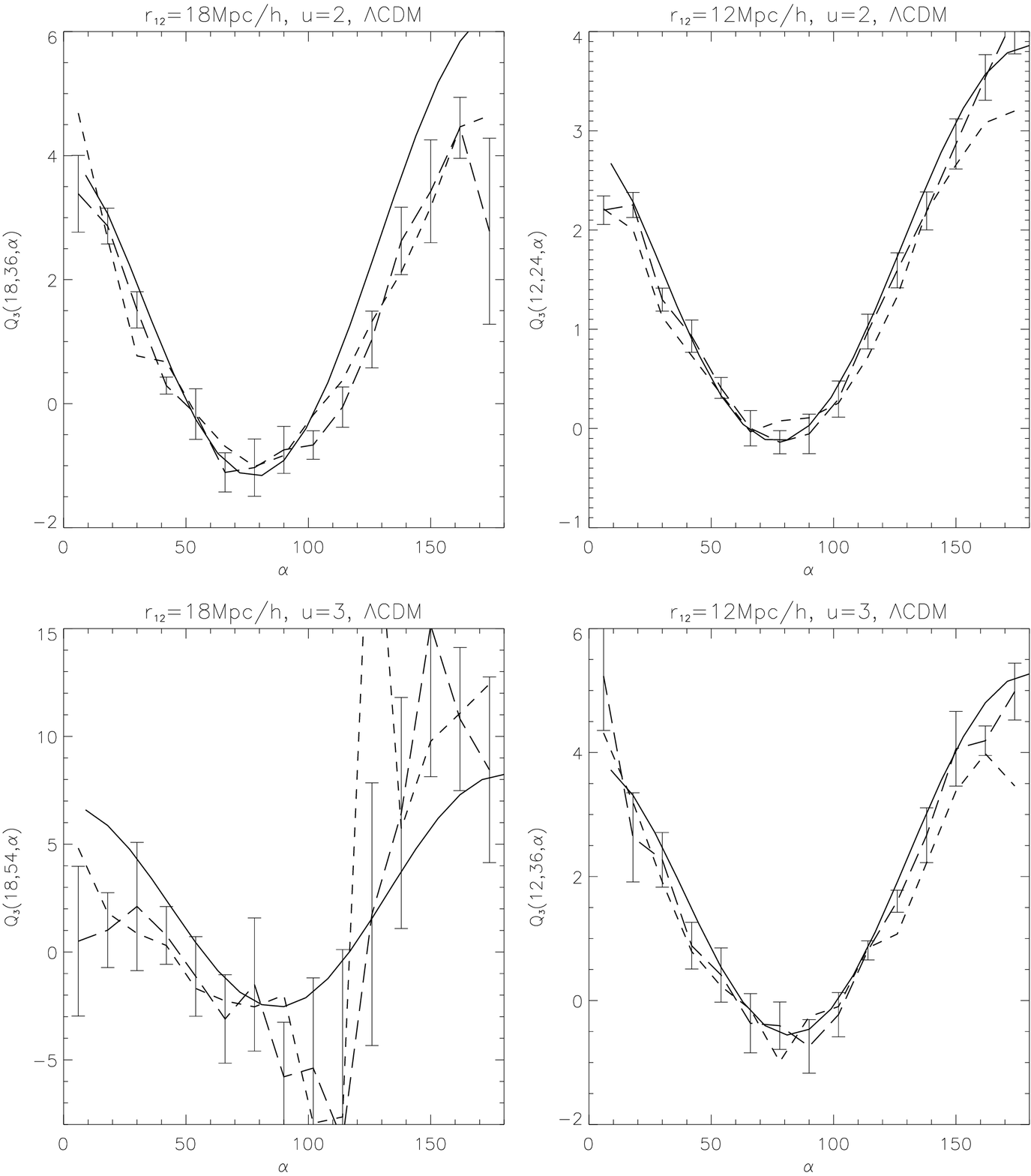}
\caption[junk]{On the left column we show results for the $\Lambda CDM$ model
with $r_{12}=18\mpc$ and on the right we show results for the same model with 
$r_{12}=12\mpc$ .The solid lines correspond to theoretical perturbation 
theory, the long-dashed lines 
correspond to the tree-level recovered
from simulation by using our method in eq. ~\ref{nintheq} and the short-dashed 
lines 
correspond to the up to 1-loop level 
$\sigma_8=1$ recovered from simulations. We only show error bars in the curve 
in which they are larger.}
\label{fig6} 
\end{figure*}

In figure ~\ref{fig6} we show plots for $Q_3$ for the $\Lambda CDM$ model with
$u=r_{23}/r_{12}=1$ (top row), $u=2$ (middle row) and $u=3$ (bottom row) for
two values of $r_{12}$: $r_{12}=18\mpc$ for the left column and
$r_{12}=12\mpc$ on the right one. The solid lines correspond to theoretical
perturbation theory, the long-dashed lines correspond to the tree-level
recovered from simulation by using our method in eq. ~\ref{nintheq}, the 
short-dashed lines correspond to the 1-loop level ($\sigma_8=1$) recovered from
simulations. Note the difference range of values of $Q_3$ in each panel.  The
purpose of this figure is to compare the effects at different scale lengths.
On the top left plot one can see how at somewhat large scales, $\alpha > 100$
deg. simulations are below the theoretical prediction. This effect is due to
volume effects (or small statistical sampling); notice that $\alpha > 100$ deg
corresponds to $r_{13}> 43\mpc$ which is above one tenth of our box side,
$378\mpc$. This type of volume effects produce a (ratio) bias in the
estimators which is not taken into account by averaging
the results over different
realisations \cite{HuGa99}.  For larger $u$ this effect is even worse. This can
be seen in the bottom left panel, where the error bars increase and the
estimation becomes quite uncertain.

At smaller scales (right panels) volume effects are smaller and there
is a general agreement of tree-level PT with simulations. Higher
order corrections are small.

\section{Conclusions}

In this paper we have studied how to obtain the 3-point correlation function
from simulations and we have compared them to the predictions of
perturbation theory (PT). 
In order to obtain the 3-point correlation function from
simulations in a reasonable amount of time we have devised a method which
dramatically reduces the number of operations to be carried out. The method,
explained in detail in section ~\ref{thealgo}, basically consists of assigning
a list of neighbours to every node of a previously gridded box. This method,
allows us to compute the $\zeta$, $\zeta_H$ and $Q_3$ for
several million particles in only a few minutes.

In section ~\ref{disen} we also propose a new method to separate from
simulations the tree-level and 1-loop contribution to the 2 and 3-point
functions. This allows a more clear comparison to the theoretical predictions.
We obtain that on sufficiently large scales, simulations results are in good
agreement with leading order perturbation theory. This seems in contradiction
with the results of ~\cite{MS94,JB97}; but the contradiction is only apparent.
In these previous analysis they focus on the results of small scales (less than
$8 \mpc$). Here we have shown that next to leading order (i.e. 1-loop)
corrections are important on these scales and that one needs to consider
larger scales (or earlier times) in order to reach the leading order
(tree-level) regime. Once this is taken into account there does not seem to be
a contradiction with what was found in~\cite{MS94,JB97}.  On even larger scales,
the 2-point correlation goes to zero and the results are dominated by sampling
errors.  The goodness of the agreement and the specific scales where PT is
valid, depend on the shape of the initial fluctuations (i.e. the $\Gamma$
parameter in CDM models), the redshift ($\sigma_8$), the particular 3-point
configuration under study and the level of accuracy in the comparison (i.e.
the sampling errors).  For the currently favoured $\Lambda CDM$ model at
$\sigma_8=1$ there is reasonable agreement at leading order for scales larger
than $20 \mpc$.

We also characterise next to leading order (1-loop) corrections and find that
in the $\Lambda CDM$ model they seem to change sign around $12 \mpc$ in $Q_3$
(see fig.\ref{q3c80dat6}).  At $12 \mpc$, loop corrections seem to cancel and
the tree-level PT prediction gives a better match to the simulations at this
scale than at slightly larger scales (this statement would depend on the
accuracy of our determination, e.g. the size of the sample, as errors increase
with scale). In section \S\ref{loop} we find that this change of sign in the
1-loop term is not clearly correlated to the critical index found in the so
called "previrialization effect"~\cite{SF96,LJBH96} for the 2-point function
and also found in the bispectrum ~\cite{S97}.

We have noted a crossing of the different shape predictions
at $\alpha \simeq 120$ degrees in both power-law and CDM
cases, even for different values of $u$. That, is when one of the
interior angle of the triplet is about
$120$ the value of $Q_3$ is  $Q_3 \simeq 1.2$, independently 
of the scales involved!. 
This seems to be a characteristic feature of the
tree-level gravitational evolution, but only holds  approximately
after loop corrections. We do not yet have a physical interpretation
for this feature, but it could be used to test gravitational
instability against observational data.

Surprisely, even though PT has been known for many years now
\cite{Peebles80,Fry84}, the results presented here are the first ones to show
a comparison of PT with the 3-point function in N-body simulations at the
appropriate scales. Most of previous comparisons of the 3-point statistics in
PT to simulations were done in Fourier space \cite{SF96,S97,S00}, in angular
space \cite{FG99} or for smoothed fields \cite{B94,BGE95,BCGS01}.  The
comparison presented here for the 3-point function (in configuration space)
extend, for the first time, to the weakly non-linear scales were this
comparison is meaningful.  Not Surprisely \cite{BCGS01}, we find an excellent
agreement with tree-level PT theory. Moreover we characterise the range of
validity of this agreement and the next order (1-loop) corrections.

\section*{Acknowledgments}

We acknowledge support by grants from IEEC/CSIC and DGI/MCT(Spain) projects
BFM2000-0810 and PB96-0925, and Acci\'on Especial ESP1998-1803-E.


\end{document}